\documentclass[12pt,preprint]{aastex}

\slugcomment{To Appear in the Astronomical Journal (April 2004)}

\shorttitle{Brightness Variations of $\eta$ Car}
\shortauthors{Martin et al.}

\begin{document}


\title{Eta Carinae's Brightness Variations Since 1998:  HST Observations of the Central  Star\altaffilmark{2,3}}

\revised{2 January 2004}

\author{J. C. Martin}
\email{martin@aps.umn.edu}
\affil{University of Minnesota Astronomy Department\\Minneapolis, MN 55455}
\author{M. D. Koppelman}
\email{michael@aps.umn.edu}
\affil{University of Minnesota Astronomy Department\\Minneapolis, MN 55455}
\and
\author{the HST Eta Carinae Treasury Project Team\altaffilmark{1}\\}

\altaffiltext{1}{This research was conducted as part of the Eta Carinae Hubble Space Telescope Treasury project via grant no. GO-9420 from the Space Telescope Science Institute.  The HST is operated by the Association of Universities for Research in Astronomy, Inc., under NASA contract NAS5-26555.}
\altaffiltext{2}{Some of the data presented in this paper were obtained from the Multi-mission Archive at the Space Telescope Science Institute (MAST). STScI is operated by the Association of Universities for Research in Astronomy, Inc., under NASA contract NAS5-26555. Support for MAST for non-HST data is provided by the NASA Office of Space Science via grant NAG5-7584 and by other grants and contracts.}
\altaffiltext{3}{Some of the data presented in this paper were obtained from the American Association of Variable Star Observers International Database.}



\begin{abstract}
We have measured the brightness variations in $\eta$ Carinae for the past six years using the Hubble Space Telescope Space Telescope Imaging Spectrograph and Advanced Camera for Surveys.  Unlike ground-based data, observations by the HST allow direct measurement of the brightness of the central star by resolving it from the surrounding bright ejecta.  We find interesting behavior during 2003 in the continuum and H$\alpha$ emission.  The data show that the established long term brightening trend of $\eta$ Car continues, including regular events which coincide with the 5.5 year spectroscopic cycle and other more rapid and unexpected variations.  In addition to the HST data, we also present ground-based data obtained from the AAVSO which show many of the same features.  The dip in the apparent brightness of the central star at the time of the 2003.5 event is wavelength dependent with no decrease in the continuum.  These observations cast doubt on a simple eclipse or occultation as the explanation for the dip and place constraints on the models for the
event.

\end{abstract}




\section{Introduction}

Two unforeseen developments in recent years have made continued 
photometry of \objectname[HD 93308]{$\eta$ Car} particularly important:  Distinctive
brightness variations in the near IR accompany its mysterious 5.5-year spectroscopic 
cycle \citep{feast01,whitelock94} and, possibly independent of the cycle \citep{damineli96}, the star brightened at a surprising rate after 1997 \citep{IAUC7146,davidson99a,vangenderen99,sterken99}.  Neither of these phenomena has been explained and the data since 1998 are seriously incomplete.  Meanwhile, the long-term brightening trend continues;  for general information about $\eta$ Car with many references, see \citet{davidson97}, and \citet{davidson00}.   


Photometry of this bright object is difficult for at least two reasons:  \\   
\begin{enumerate}
\item{At visual wavelengths, normal ground-based observations 
represent mainly the surrounding ``Homunculus'' ejecta-nebula, which 
appears much brighter than the central star and has structure 
at all radii from 0.2 to 8 arcseconds.  So far the only available measurements
of just the central star have been made with the Hubble Space Telescope
(HST).  Although the Homunculus is primarily a reflection nebula, 
its apparent brightness measured with respect to the central star has changed greatly.   
During 1998-99, while ground-based observations showed about a 
0.3-magnitude brightening of Homunculus plus star, the star itself 
nearly tripled in apparent brightness \citep{davidson99a}! This discrepancy presumably 
involves dust along our line of sight to the star, but, as mentioned 
above, it is only vaguely understood at this time. }
\item{Numerous strong emission lines produced in the stellar   
wind perturb the results for standard photometric systems.   
H$\alpha$ and H$\beta$ emission, for example, have equivalent widths 
of about 800 and 180 \mbox{\AA} respectively in HST spectra of $\eta$ Car.  
Therefore, broad-band $U$, $B$, $R$, and $I$ magnitudes,
and most medium-band systems as well, are poorly defined for this 
object.  Photometry around 5500 \mbox{\AA}, e.g. broad-band $V$, is relatively 
free of strong emission lines, but transformations from instrumental 
magnitudes to a standard system are imprecise because they involve
the other filters (\citet{davidson99a}, \citet{sterken01}, \citet{vangenderen03}, and references therein).}
\end{enumerate}

In this paper we report photometry {\it of the central star\/} obtained with two HST instruments since 1998.  For reasons noted above, these observations are quite distinct from all ground-based data; and the star's photometric behavior during 2003 proved to be especially interesting.  First we employ acquisition images produced 
by the Space Telescope Imaging Spectrograph (STIS), because these 
are numerous and internally reliable;   $\eta$ Car has been observed
many times with this instrument since 1998, especially during 
mid-2003 when the most recent ``spectroscopic event'' occurred.    
These broad-band red data cannot be transformed to any standard
photometric system, but they are quite reliable for showing
relative fluctuations by the star (strictly speaking, the stellar
wind), uncontaminated by light from the surrounding Homunculus.  
We also report surprising variations in the star's H$\alpha$ emission brightness accompanied by measurements of the nearby continuum in the STIS slit-spectroscopy data.


We supplement the STIS data with similar measurements of a few images 
obtained in 2002--03 with HST's Advanced Camera for Surveys (ACS).  
These represent a different set of wavelength bands.    
Altogether these STIS and ACS results probably represent 
most of the photometry of $\eta$ Car that will ever be obtained 
with the HST.  Very few additional STIS observations
are expected now that the 2003 event has passed.  
Earlier HST/FOS, FOC, WFC-PC, and WFPC2 data were not obtained often enough to 
be suitable for our purposes, and in most cases would be more 
difficult to compare photometrically, for technical reasons.  We report the STIS and ACS data now, rather than waiting for two or three more data points in 2004, because the behavior during 2003 was rather unexpected and therefore deserves to be noted promptly.


In addition, we briefly present a convenient summary of $\eta$ Car's 
photometric record in the AAVSO archives.  These `V' magnitudes 
represent, of course, the entire Homunculus plus star but they 
are valuable for comparisons with the HST data and also with
ground-based photometry using different filter systems, reported 
by, e.g., \citet{feast01}, \citet{vangenderen99}, \citet{sterken96}, \citet{whitelock94}, \citet{IBVS5477}, and \citet{vangenderen03a}.


In Section 2 we present the photometry of the central star based on HST/STIS acquisition images and the ACS/HRC observations.  This is followed by a section on the ground-based data from the AAVSO and then a general discussion about how these results relate to the models which have been presented to describe $\eta$ Carinae.

\section{Space-based Photometry of the Central Star}

Normal ground-based imaging and photometry cannot resolve the central star and small structures in the Homunculus nebula.  In this study, we have used the HST/STIS acquisition images which have a resolution of about 0.1 arcseconds plus the recent ACS/HRC images which have a resolution of 0.05 arcseconds to separate the contribution of central star from the surrounding nebulously.

\subsection{HST STIS Acquisition Images}

Each set of STIS observations of $\eta$ Carinae has included a pair of acquisition images.  These images are 100x100 pixel sub-frames (5 arcseconds square) centered on the middle row and column of the CCD \citep{clampin96,downes97,kimquijano03}.  Each pair of acquisition images includes an initial targeting image and a post-acquisition image.  
Only the post-acquisition images were used to measure the brightness of the central star because the target is rarely well centered in the initial targeting images, which raises concerns about ghost images and internal reflections that may unpredictably affect them (see section 2.1.2).

The majority of the STIS acquisition images of $\eta$ Carinae have been taken using the F25ND3 filter, which is a neutral density filter that covers the wavelengths from 2000\mbox{\AA} to 11000\mbox{\AA}.  Acquisition images using other filters are ignored since they lack significant temporal baseline or coverage and there is little hope of transforming between filter functions for an object as peculiar as $\eta$ Carinae.  The F25ND3 filter includes several prominent features in the spectrum of $\eta$ Carinae (Figure 1) and changes in them contribute to any measured brightness variations.

\subsubsection{Image Processing and Bias Level Correction}

We processed the STIS acquisition images using the IRAF\footnote{IRAF is distributed by the National Optical Astronomy Observatories, which are operated by the Association of Universities for Research in Astronomy, Inc., under cooperative agreement with the National Science Foundation.} procedure stas.hst\_calib.stis.basic2d with version 3.0 of STDAS and version 2.13 of CALSTIS.  En lieu of a bias over-scan region, each STIS acquisition image has the uniform value of 1510 counts subtracted from each pixel as an approximate bias level correction for the target acquisition process.  A small uncorrected bias level remains in the acquisition images after this initial correction, so we compensate for this tiny effect by modeling it as a function of time and CCD housing temperature.  

This residual bias was measured in acquisition images of the standard star \objectname[AG+81 266]{AGK +81 266} after the normal bias correction and dark current were subtracted.  The exposures of AGK +81 266 are short (2.1 seconds) and it is more than 60 degrees above the ecliptic, so zodiacal light and other sources make a negligible contribution to the background \citep{kimquijano03}.  We found that the residual bias level in the images of AGK +81 266 (Figure 2) is roughly linear with respect to time prior to the failure of the CCD Side-1 controller on May 16, 2001 \citep{proffitt02,davis01}.  There may be higher order terms in this relation however they are not significant in this study since they effect the measurements of the central star by less than 0.001 magnitude.  Subsequent to the failure of the Side-1 controller the STIS CCD temperature regulator has been held at a constant voltage permitting the CCD temperature to fluctuate.  As a result, after the Side-1 failure the residual bias level of the observations depends on both time and CCD housing temperature.  Using the measurements of the residual bias level in the AGK +81 266 observations, the following relations were obtained for the bias level correction as a function of modified Julian Date (MJD) and CCD housing temperature (T):\\
MJD $\leqq$ 52045.5 :
\begin{equation}
BLEV = -269.84 - 0.0053*MJD
\end{equation}
MJD $>$ 52045.5 : 
\begin{equation}
BLEV = -293.34 - 0.0053*MJD - 1.51*T
\end{equation}
When the modeled bias level is subtracted from the AGK +81 266 data the background levels average to $-0.03\pm0.63$ counts/pixel/sec (Figure 2).  This correction was applied to the acquisition images for $\eta$ Carinae in place of the bias over-scan correction normally implemented in the IRAF basic2d procedure.  Typically the bias correction amounts to less than 0.005 magnitude for the measurements reported below.

\subsubsection{Aperture Photometry}

The flux from the central star was measured in each image by summing detector counts within a small circular virtual aperture. In order to avoid pixelization effects, the counts were weighted by a function of the distance of a pixel from the aperture center ($r$) which decreases smoothly to zero at the aperture radius $R$:\\
\begin{equation}
r \leqq R : f(r)=1-\frac{r^2}{R^2}
\end{equation}
\begin{equation}
r > R : f(r)=0
\end{equation}
The centering on the star is determined by a peak-up algorithm which locates the position where the sum inside the virtual aperture is maximized.  For the STIS acquisition images we used $R=3$ pixels which corresponds to a diameter of 0.3 arcseconds and includes slightly fewer than 30 pixels.  

We did not attempt to subtract a background from the measured aperture flux, but this has little effect on the results since the star provides thousands of counts/pixel/second, far greater than any other contribution within the measuring aperture due to scattered and zodiacal light which contribute only about 0.2 counts/pixel/second or less \citep{kimquijano03}.   There is no feasible way to account for or remove the contribution from diffuse circumstellar material within 0.1 arcseconds of the star.  However, STIS spectroscopic data indicate the level of contamination from this source is much smaller than the other uncertainties.

Ghost images exist in the STIS optics but do not contribute significantly to our measurements.  When a bright source is centered on the CCD, a ghost image with an peak flux of a few percent relative to the PSF maximum appears approximately 0.3 arcseconds (6 pixels) to the right (increasing column number) of the peak of the point spread function.  The 0.3 arcsecond diameter aperture used to measure the central star excludes this ghost.  However, any future work that may attempt to measure the flux from the nebula immediately surrounding the central star in these images must confront this problem.  

The reduction procedure was tested using 80 pairs of acquisition images of the flux standard AGK +81 266 (Figure 3).  The measured aperture photometry for this standard shows a constant level with an r.m.s scatter of $\pm0.037$ magnitudes.  It should be noted that AGK +81 266 was observed using the F28x50LP ``long-pass'' filter rather than the F25ND3 filter which was used for the $\eta$ Carinae observations.  At the same time, the gross spectral energy distribution of AGK +81 266 (spectral class B5) and $\eta$ Carinae are very dissimilar.  This is not significant since AGK +81 266 is only used to demonstrate that this method accurately measures a constant brightness over the temporal baseline for an established flux standard.  The brightness of $\eta$ Carinae is {\em not\/}  measured relative to AGK +81 266.


\subsubsection{Results}

In Figure 4 and Table 2 we present our results for the STIS acquisition images as instrumental magnitudes calibrated for long term sensitivity effects using the photometric constants calculated by the reduction pipeline \citep{kimquijano03} and scaled relative to the value on 1999 February 20 (1999.14).  These represent an extremely broad-band wavelength sample (Figure 1). For the purpose of comparison with familiar photometric systems,    Table 1 gives approximate Johnson $V$ \citep{johnson53} and Cousins $R$ and $I$ \citep{cousins76} magnitudes at the reference epoch (1999.14).  These were calculated using flux-calibrated STIS spectroscopic data \citep{davidson03}, a flux-calibrated spectrum of Vega \citep{colina96}, and the appropriate response functions \citep{johnson51,bingham74}.  For reasons noted in Section 1 and uncertainties in the STIS slit throughput for an object which is not exactly a point source, entries for $\eta$ Carinae in Table 1 probably have uncertainties of roughly $\pm0.05$ magnitudes and could possibly be somewhat worse.  It is also possible that the photometric color may trend mildly blue-ward as the object brightens \citep{davidson99a}. 

The Homunculus is currently of 5th magnitude according to {\it ground-based\/} Johnson $V$ photometry (Section 3 below), but the central star is fainter than $V$ = 7.0 as stated in Table 1.  Its apparent continuum is red, while stellar-wind emission lines           contribute a significant fraction of its brightness.  In the 1999.14 spectrum, H$\alpha$ and other emission features contribute about 16\% of the flux in $V$ (4\% from H$\alpha$) and 38\% of the flux in $R$ (35\% from H$\alpha$).  Meanwhile, there is about a 5\% affect on the $I$ flux from the Hydrogen Paschen lines which are in emission.  Because these emission features are constantly changing, it is not possible to formulate a consistent set of transformations to convert the STIS magnitudes to more familar Johnson $V$ and Cousins $R$ and $I$ magnitudes for $\eta$ Carinae.  However, the values presented for 1999.14 can serve as a rough zero point.

Figure 4 shows that the dramatic brightening reported by \citet{davidson99a} leveled off by 2000, but may have resumed at least temporarily around the mid-2003 spectroscopic event.  The dip in brightness near the time of the event (Figure 6) may be reminiscent of the near-infrared  behavior reported by \citet{feast01}.  This will be discussed further in Section 4.

\subsubsection{Concerning H$\alpha$ Emission}

Eta Carinae's broad H$\alpha$ emission, produced in the stellar  wind at radii of several AU, is tremendously strong  (equivalent width usually about 800 \mbox{\AA}).   This feature did not vary much relative to the continuum in STIS data from 1998 to 2002,  but recently it has faded to a surprising degree as reported below.  Since H$\alpha$ contributes appreciably to the STIS acquisition images, we have investigated the effects on our photometry.

Davidson (2003) has measured the equivalent width of H$\alpha$ and estimated the absolute level of the nearby continuum in short-exposure STIS slit data.  H$\alpha$ was integrated from 6509 to 6650 {\AA}, and the continuum flux was averaged between 6740 and 6800 {\AA} (an interval that is practically uncontaminated by emission lines).  These continuum measurements have the advantage of being approximately calibrated in physical units, but on the other hand may be underestimated if the slit was not perfectly centered on the star.  Fortunately, the H$\alpha$ equivalent width measurements are quite insensitive to slit position.  The results are shown in Figure 5.  We estimate that H$\alpha$ contributes from 8\% to 16\% of the STIS/F25ND3 counts, varying with time as listed in Table 2.  Open triangles in Figure 4 represent the instrumental magnitudes that would have been observed if H$\alpha$ were not present;  evidently the brightness minimum in early 2003 and the subsequent rapid brightening are not due to this emission line.   The measured brightness fluctuations in the central star appear to be mainly (though not entirely) changes in the continuum brightness. 

In the six months preceding the 2003.5 event (December 2002 to June 2003), the equivalent width of H$\alpha$ declined by a factor of nearly 2, while its line profile evolved in an interesting and unexpected way which is beyond the scope of this paper.  The other Balmer lines behaved in a similar fashion.  Moreover, in July and September 2003 they did not return to the state observed by STIS in 1998 at the corresponding point in $\eta$ Car's 5.54-year cycle.  A more in-depth account of these effects will be reported in a paper which is now in preparation \citep{davidson04}.

\subsection{ACS/HRC Observations}

HST ACS/HRS observations of $\eta$ Car were obtained for the HST Treasury Project beginning in October 2002.  Bias corrected, dark subtracted, and flat fielded ACS/HRC images were obtained from Space Telescope Science Institute via the Multi-mission Archive 
(MAST)\footnote{http://archive.stsci.edu/}.  Observations which had overexposed the central star were not used and no drizzle or dithering correction is applied.  

The ACS/HRC images were taken in four filters which cover near UV and optical wavelengths (Figure 1).  Each of these filters is influenced by different types of spectral features.  The UV filters (HRC/F220W \& HRC/F250W) are heavily influenced by the ``Fe II forest'' \citep{cassatella79,altamore86,viotti89}.  The opacity of this ``forest'' is known to dramatically increase during an event \citep{davidson99,gull00}.  The HRC/F330W filter is sensitive to both the Balmer continuum and a number of emission features while the HRC/F550M filter, strategically placed between H$\alpha$ and H$\beta$, almost exclusively measures the continuum.  

The relative brightness of the central star in the ACS/HRC images is measured with the same weighted 0.3 arcsecond diameter aperture used to measure the STIS acquisition images.  On the scale of the ACS/HRC this corresponds to $R=5$ pixels and includes about 79 pixels inside the aperture.  The results are given in Table 3 and Figure 6.  The magnitudes given are on the STMAG system \citep{holtzman95} as calibrated by the photometric keywords calculated by the reduction pipeline \citep{pavlovsky03}.  Note that the HRC/F220W and HRC/F250W curves show a significant dip at the time of the 2003.5 event, due to  the \ion{Fe}{2} absorption as noted above.  Interestingly, the visual continuum measured by the HRC/F550M filter does {\it not\/} show the decrease in brightness seen at longer wavelengths in the STIS data.

\section{Ground-based AAVSO Data}

While space-based photometry is able to resolve the central star, ground based photometry measures the integrated brightness of both the star and the Homunculus nebula.  However, ground-based photometry is still valuable because it has a longer temporal baseline and shows many of the same brightness variations, albeit at different amplitudes.  Also as more publicly funded small telescopes are closed the data from the AAVSO and groups like La Plata observatory \citep{IBVS5477} will become increasingly useful for monitoring objects like $\eta$ Car.

One of the few sources for long term monitoring of $\eta$ Carinae is the
American Association of Variable Star Observers (AAVSO)
\citep{mattei98}. From 1968 through mid-2003 there are 6640 observations from 110 observers
situated around the globe\footnote{http://www.aavso.org/news/etacar.shtml}. 
These data are from individual observers who have determined the brightness of $\eta$ Carinae by visually comparing it with the brightness of other stars.  Ten observers account for 69\% of these observations while 36\% of the observations are from the two most prolific observers:  M. Daniel Overbeek \citep{mattei03} and Albert F. Jones. On average, there are 60 observations per observer but the median is 9 observations per observer.

We averaged the observations in 90 day bins with a median of 32 observations in each bin. The bottom panel of Figure 7 shows binned averages from 1968 through mid 2003 with a typical standard deviation of 0.2 magnitudes, which is consistent with the accepted usual r.m.s. error of 0.1 -- 0.2 magnitudes for observations of this type.  The top panel of Figure 7 shows the photoelectric Johnson $V$ band photometry (PEPV) from Stan Walker \citep{davidson99a}, photoelectric observations from R. Winfield Jones obtained through the AAVSO, and the average $V_J$ magnitude from ground based observations reported  in Figure 2 of \citet{davidson99a}.  The data in both panels follow the same general trend and compare well with the data presented in Figure 2 of \citet{sterken96}.  Even the most primitive AAVSO data track the professional photometry well enough to show irregular behavior.  The maximum noted by \citet{sterken96} at 1982.11 is visible in all the data, though the errors in the binned AAVSO data make it harder to recognize.  Each dataset also shows an overall brightening trend in $\eta$ Carinae over the past thirty years along with a notable jump in brightness in 1998-1999 \citep{davidson99a}.

The binned AAVSO data for the last five years (Figure 8) also follows a trend similar to the brightness variations in the central star measured from the STIS acquisition images (Figure 4).  Of particular note, both datasets show the brightening of $\eta$ Carinae peaked in 2000, fell off a bit, and then peaked again in early 2002 before brightening through the 2003.5 event.  

\section{Analysis \& Conclusions}

The data presented here from the HST and ground-based observations confirm the dramatic brightening in both the nebula and central star of $\eta$ Carinae which was previously reported by \citet{IAUC7146,davidson99a} and \citet{sterken99} along with a more gradual brightening spanning the past thirty years.  All the available data also show that the apparent brightness of $\eta$ Carinae fluctuates with no obvious period.  We also see intervals (on the order of months) with short lived increases in apparent brightness similar to those reported by \cite{vangenderen03}.  The 2003.5 event, as observed in the brightness fluctuations of the central star at visual wavelengths, was characterized by a roughly six month rise, then a sharp dip in most filters, followed by a recovery to the highest flux observed for the central star in recent years, which may be the start of another significant brightening episode.  Again we emphasize that since 1997 {\it the central star has brightened substantially faster than the Homunculus\/} (compare Figures 4 and 7).

The STIS, ACS, H$\alpha$, and AAVSO data all have notable trends which began about 200 days before the 2003.54 event.  The visual band photometry from the STIS, ACS, and AAVSO began to brighten significantly in January 2003 (Figures 9 \& 6 and also \citet{IBVS5477}) which coincided with a steep drop in H$\alpha$ luminosity (Figure 5) and the beginning of a dip in brightness in the near-UV (Figure 6).  The start time for this pre-event activity is important because most models which invoke a binary companion to trigger the events place the secondary on a very eccentric orbit where it is in close proximity to the primary for only a very brief period of time during each cycle.  \cite{feast01} showed that the $J$, $H$, $K$ and $L$ bands brightened roughly 500 days before the time of the events over the past two decades with an additional sharp rise in the $J$, $H$, and $K$ bands about 80 days prior to the 1998.0 event.  The X-ray flux from $\eta$ Car (\citet{ishibashi99} and M. Corcoran's web page\footnote{http://lheawww.gsfc.nasa.gov/users/corcoran/eta\_car/etacar\_rxte\_lightcurve/} for data on the recent event) also increased before the 1998.0 and 2003.54 events, preceding them by about 350 and 700 days respectively.  It is not necessary or expected that all of the pre-event activity will temporally coincide since each of these data measure activity which occurs at different radii from the star and a shock front which pushes deeper as the event proceeds will not affect each process simultaneously.  Also, as noted below there is compelling evidence that the state and distribution of the material immediately surrounding the central star has evolved over time so while the activity prior to the event should maintain roughly the same sequential order, the exact start times relative to the event could potentially change from cycle to cycle.

The ACS/HRC data show that the brief dip in apparent brightness during the 2003.5 event is not achromatic.  The ACS data show that the depth of the dip is wavelength dependent; the dip is deeper at shorter wavelengths and the brightness continually increased in the HRC/F550M filter through the last data point.  However, \citet{feast01} showed that the brief decrease or dip in luminosity occurred at all of the near-IR wavelengths during the 1998.0 event (Figure 9, reproduced from \citet{feast01}) and \citet{IAUC8160} reported similar activity during the 2003.5 event.  This leaves two possibilities which are not mutually exclusive:
\begin{enumerate}
\item{The decrease in brightness occurs later at redder wavelengths and/or}
\item{The decrease in brightness is driven entirely by increased line opacities, decreased emission lines, and Balmer and Paschen continua formed in the wind,  rather than a change in the star's continuum flux.}
\end{enumerate}

\citet{feast01} show that during the 1998.0 event, the dip in the $J$ band occurred about 35 days later than the dip in the X-Rays.  They also note that the epoch of the minimum in the dip increases slightly with increasing effective wavelength (Figure 9).  \citet{IAUC8160} reported that during the 2003.5 event the dip in the $J$ and $L$ bands began sometime between MJD 52809.5 (2003.467) and 52814.5 (2003.481), which preceded the observed dip in Johnson-Cousins $BVRI$ by about one to two weeks \citep{IBVS5477,vangenderen03a}.  (Note that wavelengths around 2 $\mu$m observed by \citet{feast01} presumably represent free-free emission in the wind at radii substantially larger than the photosphere at visual wavelengths.) Given only the HST data, one might suspect that dip in the HRC/F550M filter may have occurred between data points and the slight rise observed at MJD 52803.1 (2003.45) is the rise in luminosity which precedes the dip.  However, the HRC/550M observation at MJD 52840.1 (2003.55) shows a {\it rise\/} in flux roughly coincident with the minimum observed in $BVRI$ by \citet{IBVS5477} and \citet{vangenderen03a}.  Therefore, it is unlikely that we missed the dip in HRC/550M due to sampling.  

Likewise, strong P-Cygni absorption is present in the hydrogen lines during the event \citep{davidson99}.  Since the hydrogen emission lines contribute significantly to the total brightness of the central star, anything that changes their profiles will change the brightness measured in a filter which they influence.  The HRC/F550M filter covers none of the prominent emission lines and it shows no dip concurrent with the dip in the other filters.  Therefore, this may indicate that the bolometric luminosity of the central star is largely unaffected by the mechanism which causes the observed decrease in brightness.

\citet{davidson99a} discuss several possible explanations for the variability in the brightness of the central star of $\eta$ Carinae.  Since the star itself radiates near the Eddington Limit, there is little room for rapid or dramatic changes in bolometric luminosity.  This is supported by the fact that the surrounding nebula, which acts as a giant calorimeter for the central star, varies in brightness on a significantly smaller scale.  If the bolometric luminosity of the central star stays roughly constant, then significant variations in its brightness may be more appropriately interpreted as changes in the density, temperature, and spatial distribution of intervening circumstellar material.  An expanding shell of ejecta from an event may be able to produce the observed light curves \citep{zanella84,davidson99,davidson99a,smith03}.  Any ejected shell should start hot with high opacities from ionized metals and then later shift these opacities to the red as the shell cools and dust forms.  Particularly at early stages, this shell may be optically thin and as such would not have a significant impact on the continuum brightness.

The data show that central star of $\eta$ Carinae continues to brighten.  We are unsure how long this trend can continue but it appears to be driven by an increase in the apparent continuum brightness which may be caused by changes in the intervening circumstellar material along the line of sight, although how or why the circumstellar material may have been been altered is not completely understood.  Continued monitoring is critical since coming out of the 2003.5 event there are indications that the central star may undergo another period of sudden dramatic brightening in the near future.  

The brief decrease in the star's brightness during the 2003.5 event was more complicated than expected and changes preceding the event in the optical and near-UV started as early as January 2003.  The wavelength dependence of the dip indicates that it cannot be caused by a simple eclipse or occultation.  If the primary star is eclipsed by the presumed secondary we 
would expect to observe some decrease at all wavelengths including the HRC/550M
band which measures the continuum. On the other hand if an eclipse of the secondary
is sufficient to cause a dip of approximately 0.5 mag in the near UV flux of the central
star, then we would expect to find some spectroscopic evidence for the secondary 
in the HST/STIS spectra during the 5.5 year cycle which is not the case.  However, the data may support a shell ejection model for the spectroscopic events.  Additional data, including spectra gathered by the STIS during the 2003.5 event, should further constrain the mechanism powering these periodic phenomena.

\section{Acknowledgments}
This research was conducted as part of the HST Treasury Project on $\eta$ Carinae via grant no. GO-9420 from the Space Telescope Science Institute.  
We acknowledge Kris Davidson, Roberta M. Humphreys, and Kazunori Ishibashi for directly assisting with the data and measurements and T.R. Gull for preparing most of the HST observing plans.  Other participants in the HST Treasury Project are D. John Hillier, Augusto Damineli, Michael Corcoran, Otmar Stahl, Kerstin Weis, Sveneric Johansson, Fred Hamann, H. Hartman, Nolan Walborn, and Manuel Bautista.

We are also grateful to the AAVSO International Database for the variable star observations contributed by observers worldwide.  We individually thank Janet A. Mattei and Elizabeth O. Waagen of the AAVSO who helped us obtain the data and Stan Walker and R. Winfield Jones who provided photoelectric photometry data.  We are very appreciative of M. Feast, and P. Whitelock for allowing us to reproduce their figures.  And we thank Matt Gray and Qian An for providing the $\eta$ Carinae project with computer technical support.




\clearpage


\begin{figure}
\figurenum{1}
\epsscale{1}
\includegraphics[angle=90,scale=0.7]{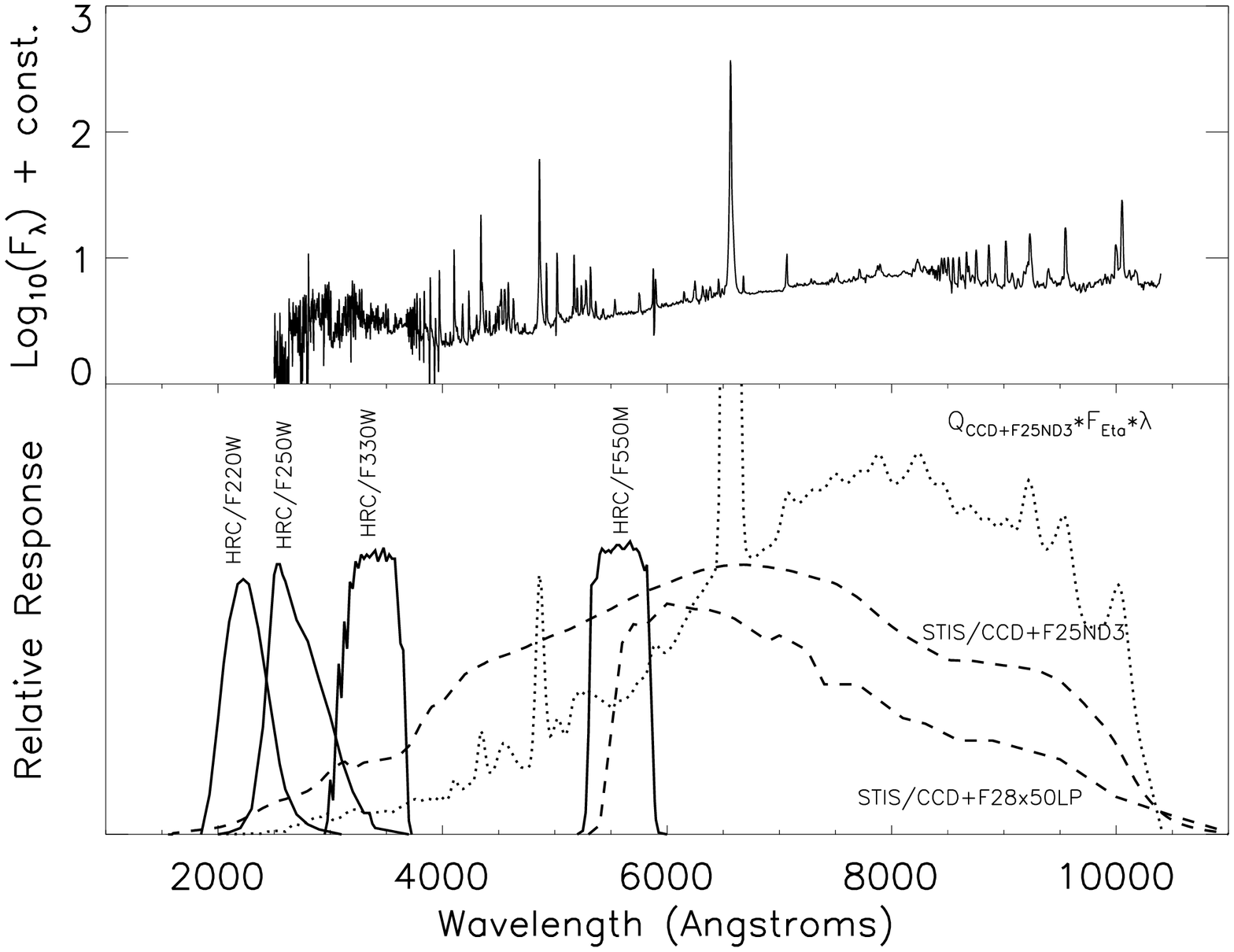}
\caption{The top panel shows the relative spectral flux from the central star of $\eta$ Carinae.  The bottom panel shows the total relative response of each CCD and filter combination used in this study on the same wavelength scale as the top panel.  For plotting purposes the curves are not representative of relative responses between filters.  STIS filters are plotted with a dashed line and ACS/HRC filters are plotted with a solid line.  The dotted line represents the product of the STIS CCD+F25ND3 response curve and the photon flux from the central star.}
\end{figure}

\begin{figure}
\figurenum{2}
\epsscale{1}
\includegraphics[angle=90,scale=0.7]{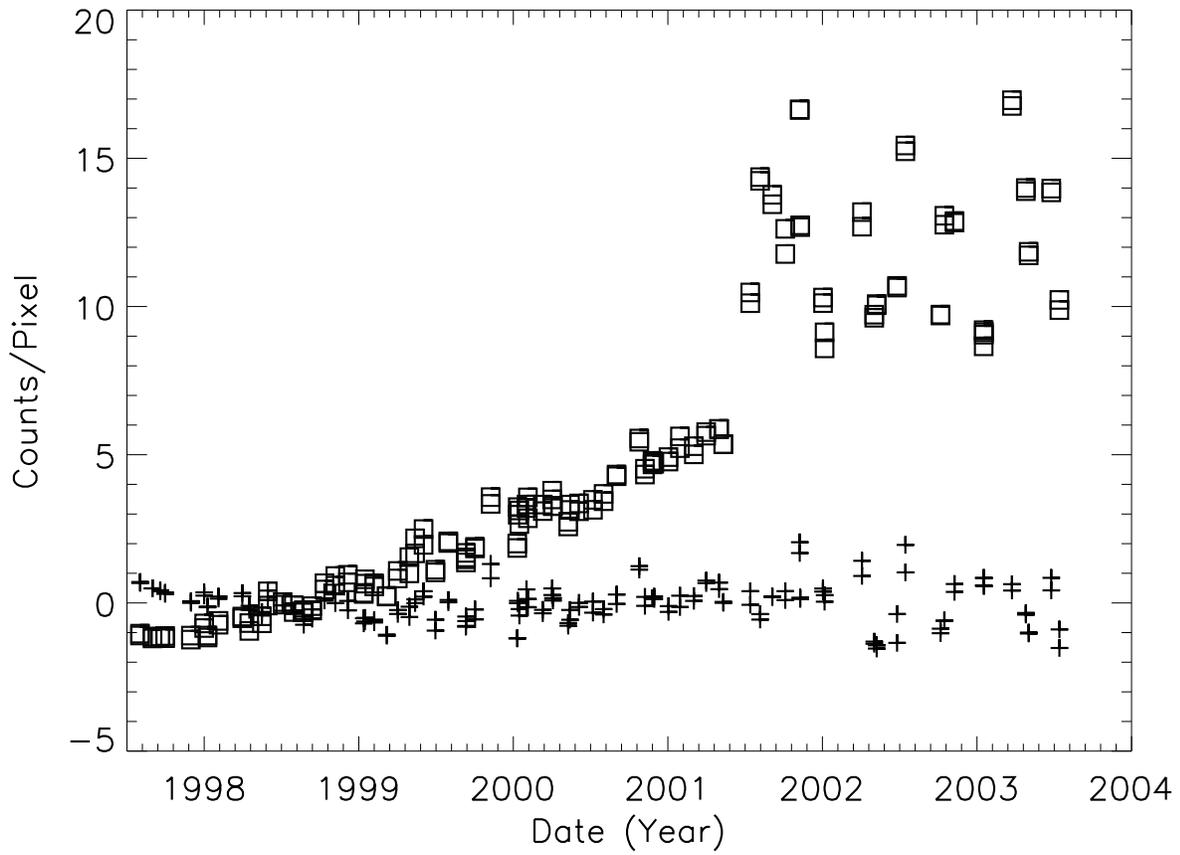}
\caption{A plot of the residual bias level in the 80 STIS acquisition image pairs of flux standard AGK +81 266.  Open squares are the residual bias level before subtracting the modeled correction.  Crosses (+) are the bias level after subtracting the modeled correction.}
\end{figure}

\begin{figure}
\epsscale{1}
\figurenum{3}
\includegraphics[angle=90,scale=0.7]{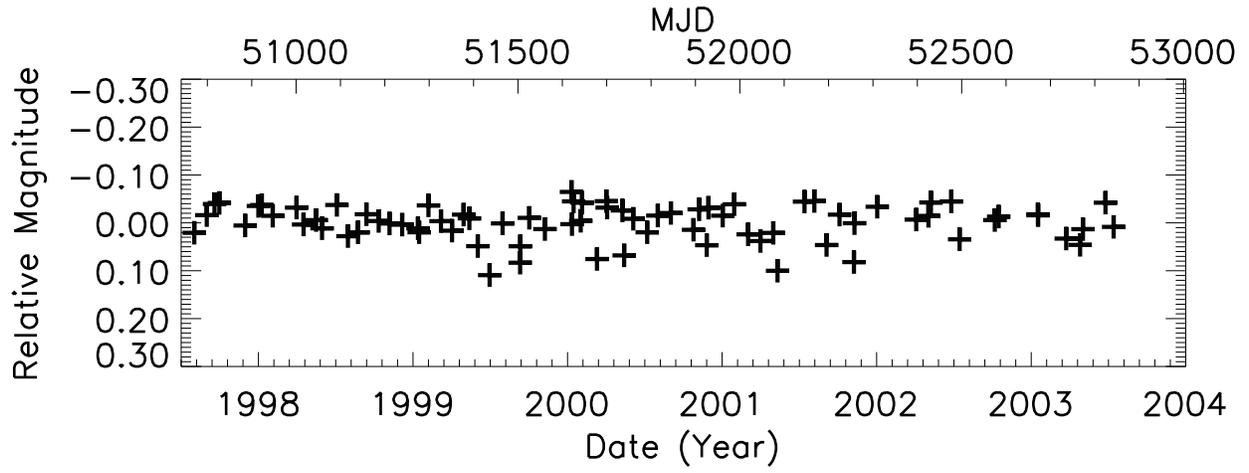}
\caption{The relative brightness of flux standard AGK +81 266 measured using a 0.3 arcsecond diameter weighted aperture.}
\end{figure}

\begin{figure}
\figurenum{4}
\epsscale{1}
\includegraphics[angle=90,scale=0.7]{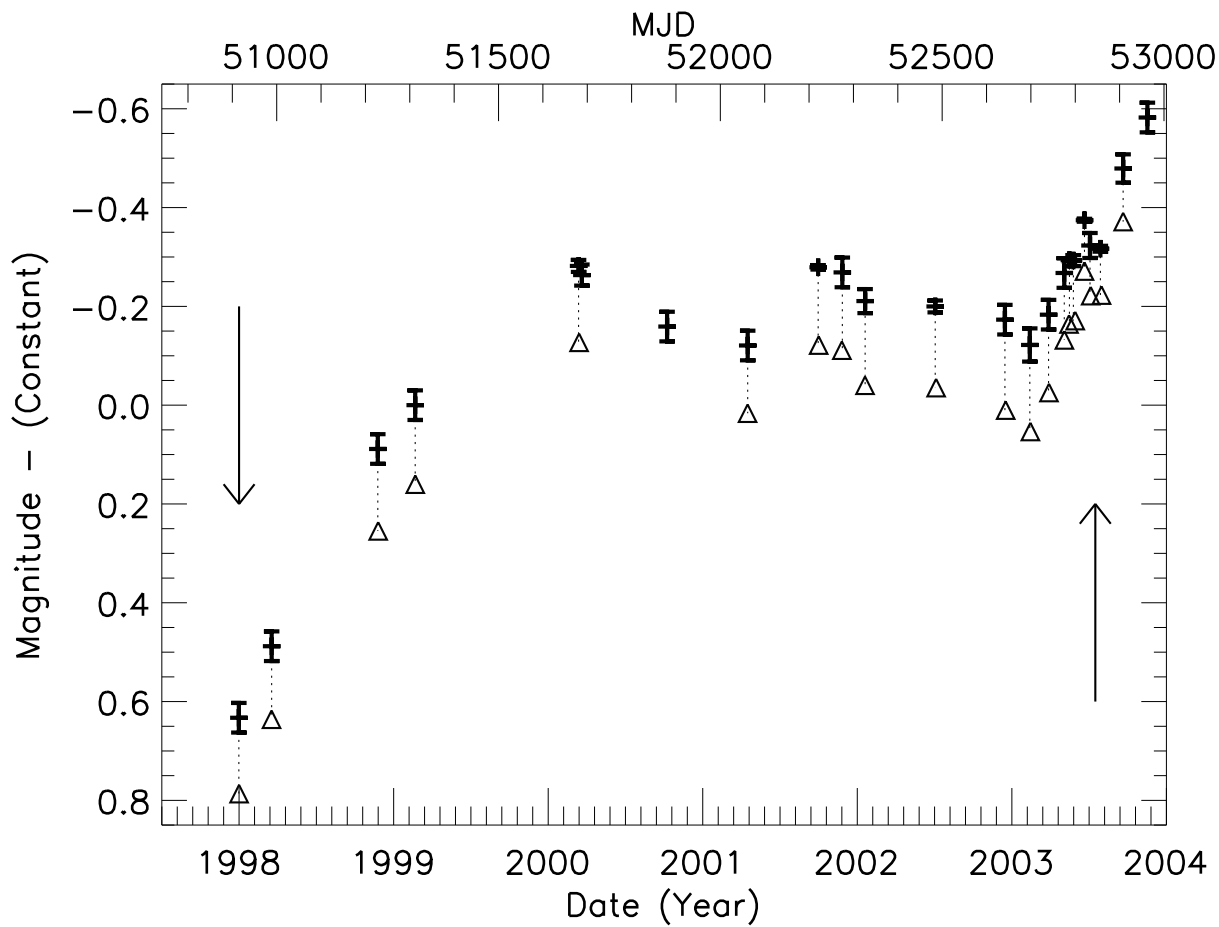}
\caption{The crosses (+) with error bars are the average relative brightness of the central star in the STIS acquisition images (Table 2).  The error bars are the one sigma standard deviation of the measurements included in each point.  The error bars plotted for points including only one measurement are the r.m.s. deviation of the brightness measured for the flux standard AGK +81 266.  The open triangles are the magnitudes with the contribution of H$\alpha$ subtracted.  Vertical arrows mark the times of the 1998.0 and 2003.54 spectroscopic events.  See Figure 6 for an expanded view of the data around the 2003.5 event.}
\end{figure}

\begin{figure}
\figurenum{5}
\epsscale{1}
\plotone{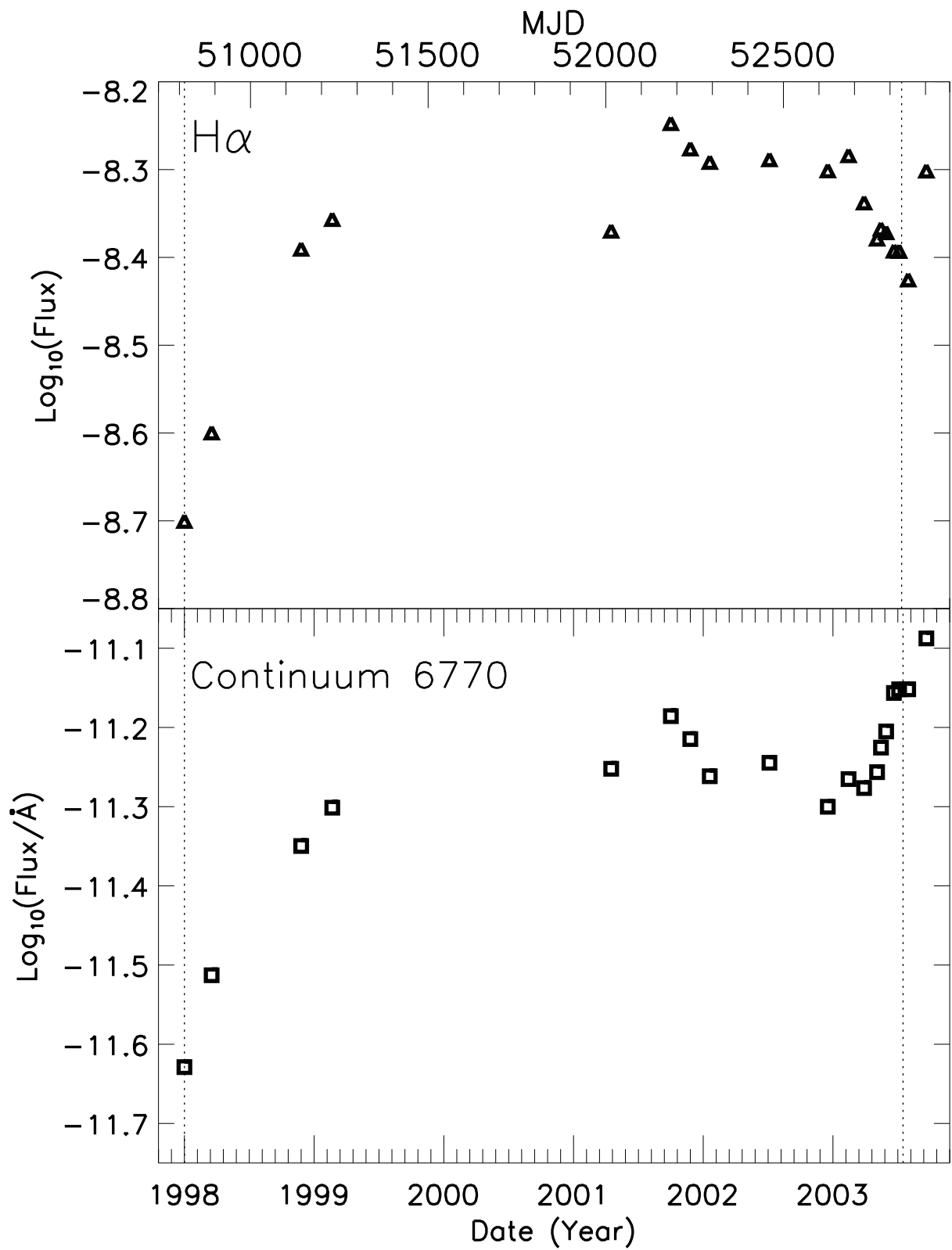}
\caption{The flux of features in the STIS spectrum of the central star.  The top panel shows the log of the total integrated flux of H$\alpha$ in erg/cm$^2$/s.  The bottom panel shows the log of the continuum flux measured at 6770 \mbox{\AA} in erg/cm$^2$/s/\mbox{\AA}.  The dotted vertical lines mark the spectroscopic events at 1998.0 and 2003.54.}
\end{figure}

\begin{figure}
\figurenum{6}
\epsscale{0.75}
\plotone{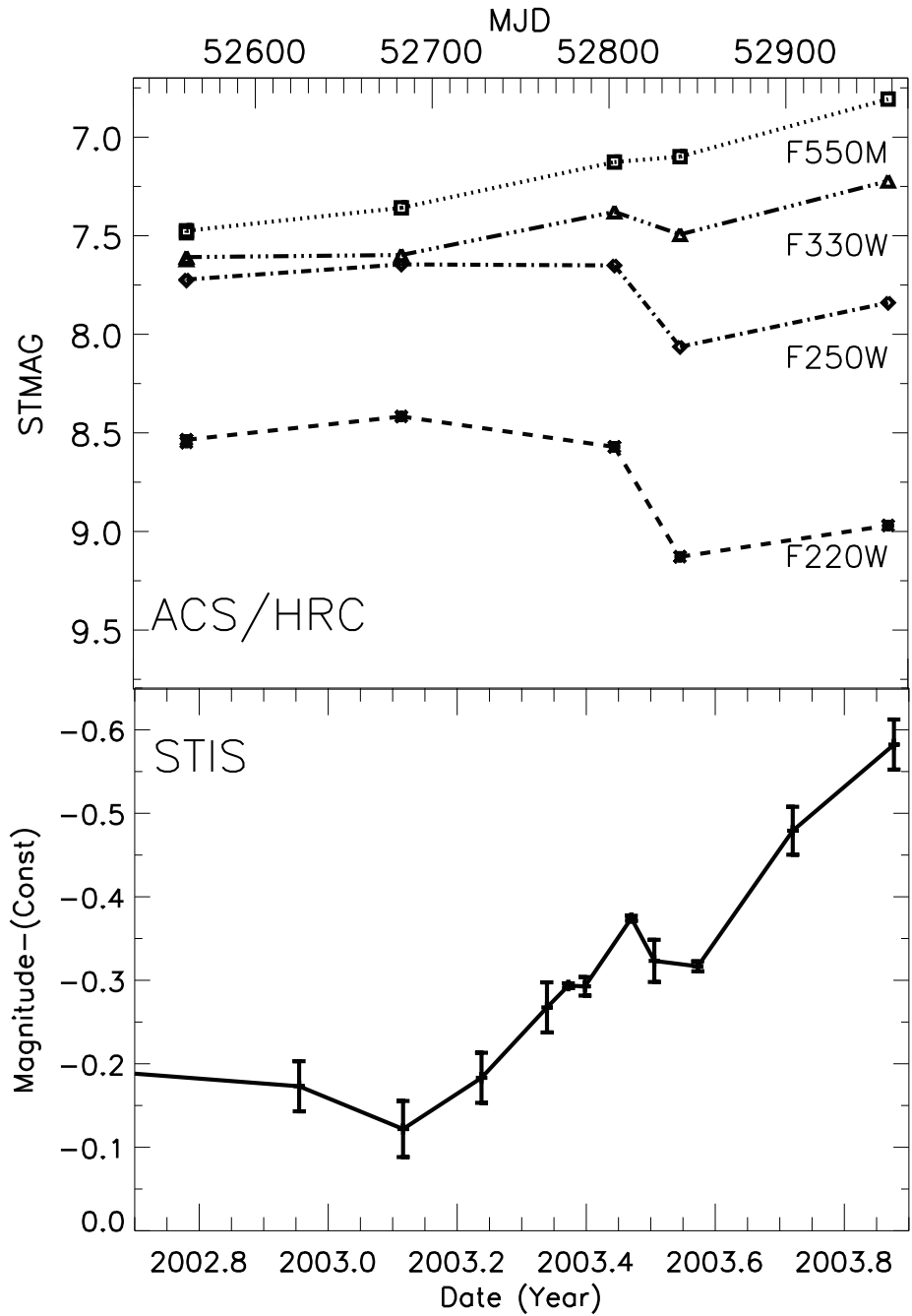}
\caption{The top panel shows the brightness of central star in the ACS/HRC images with each point representing the average from several individual exposures (Table 3).  The data for each filter is represented as its own line, labeled on the right.  The bottom panel shows the data from the STIS acquisition images (solid line).}
\end{figure}

\begin{figure}
\figurenum{7}
\epsscale{.7}
\plotone{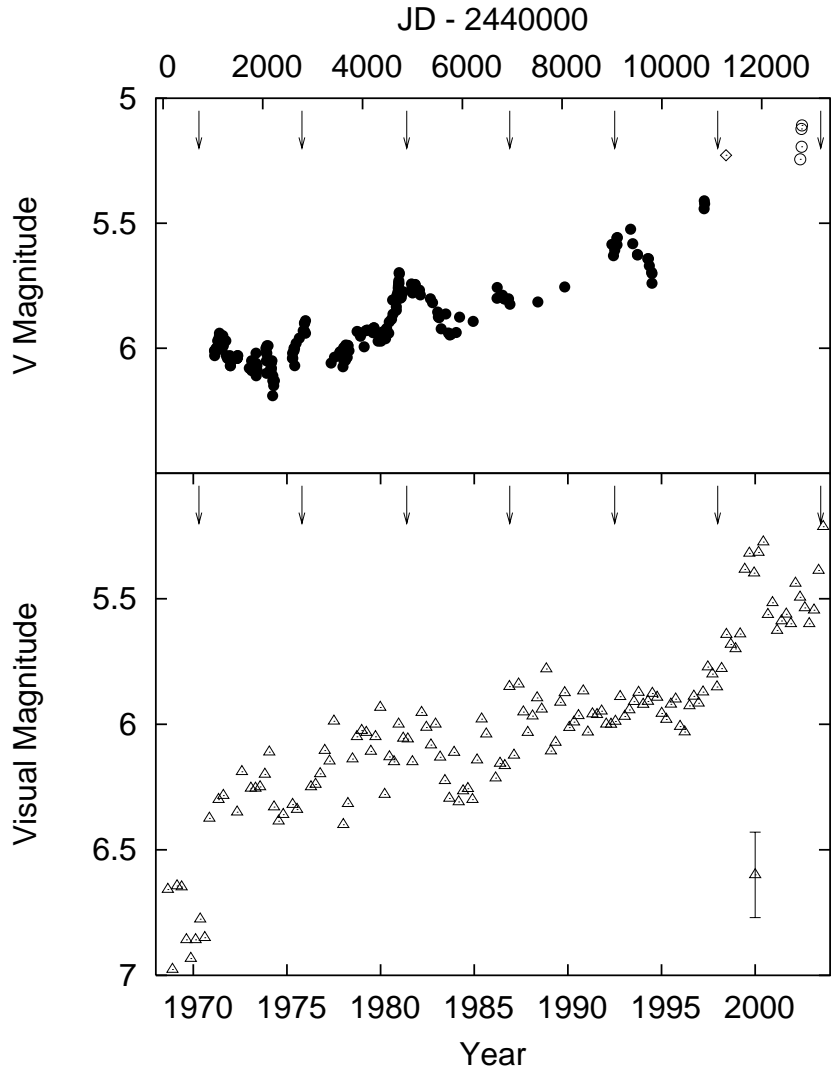}
\caption{The light curve of $\eta$ Carinae from ground based observations. The top graph shows the photoelectric $V$ magnitudes (PEPV) from Stan Walker (filled circles), photoelectric V observations by R. Winfield Jones obtained from the AAVSO (open circles), and the average $V_J$ magnitude from the ground based observations reported by \citet{davidson99} (open diamond).  The bottom panel shows the averages of the visual AAVSO observations in bins of 90 days. A representative error bar for the binned AAVSO observations is shown in the lower right; see also Fig.\ 8.  The arrows along the top of each panel mark the interval of past events with the 5.54-year period adopted in 2001 for the Treasury Project proposal.  This period was based on photometric details reported by \citet{feast01} and various other data, and it predicted the timing of the 2003 event quite well.}
\end{figure}

\begin{figure}
\figurenum{8}
\epsscale{1}
\includegraphics[angle=270,scale=0.7]{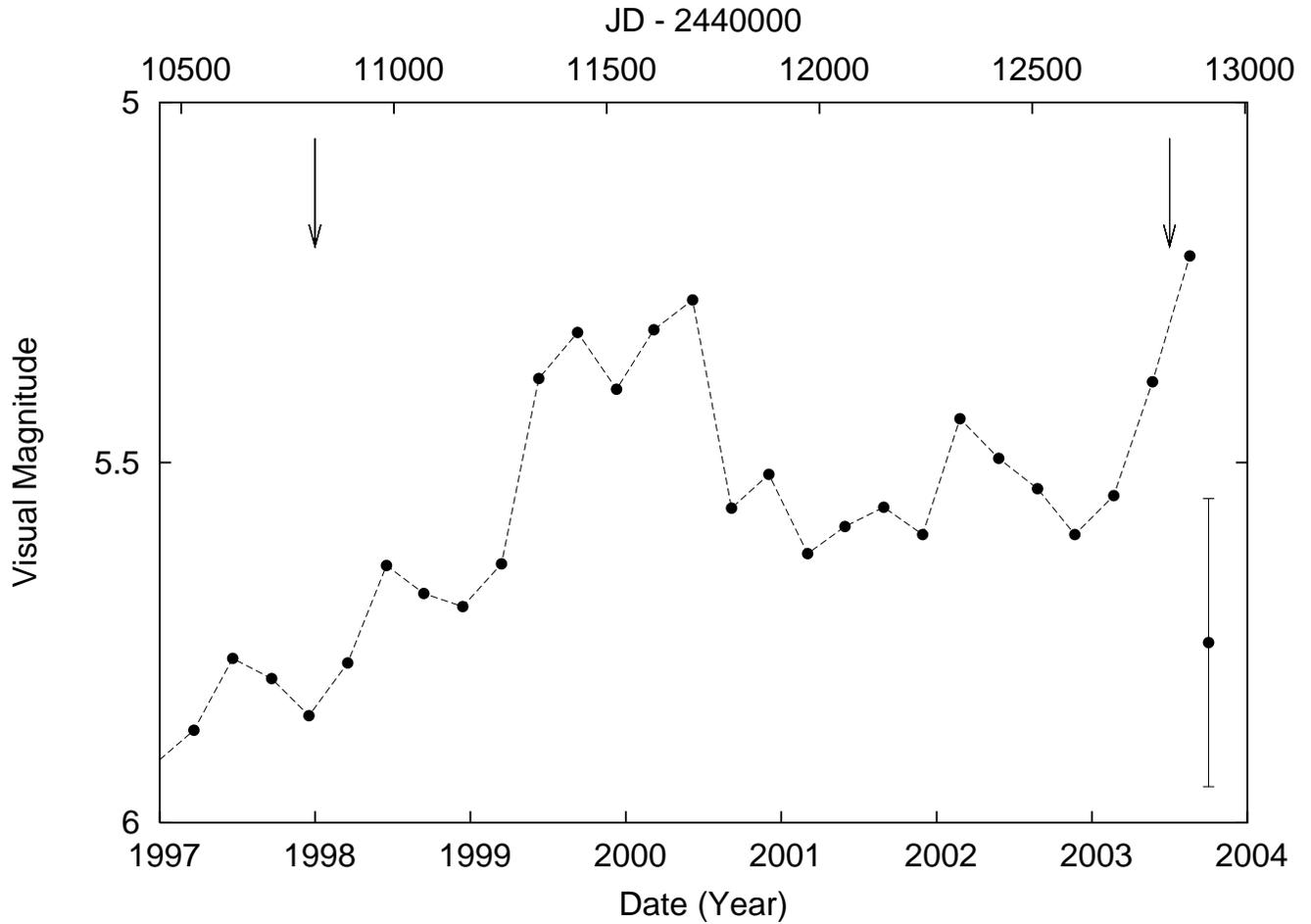}
\caption{Eta Car light curve from 1997 through 2003 of the visual AAVSO observations averaged in bins of 90 days. A representative error bar for a single observation is shown in the lower right; each point here represents typically about 32 observations.  The arrows on the top of the plot mark the 1998.0 and 2003.54 spectroscopic events.  Formal error estimates are not fully justified in this case, but the r.m.s. uncertainty for each plotted point is probably about 20\% as large as the single-observation error bar.}
\end{figure}

\begin{figure}
\figurenum{9}
\epsscale{0.75}
\plotone{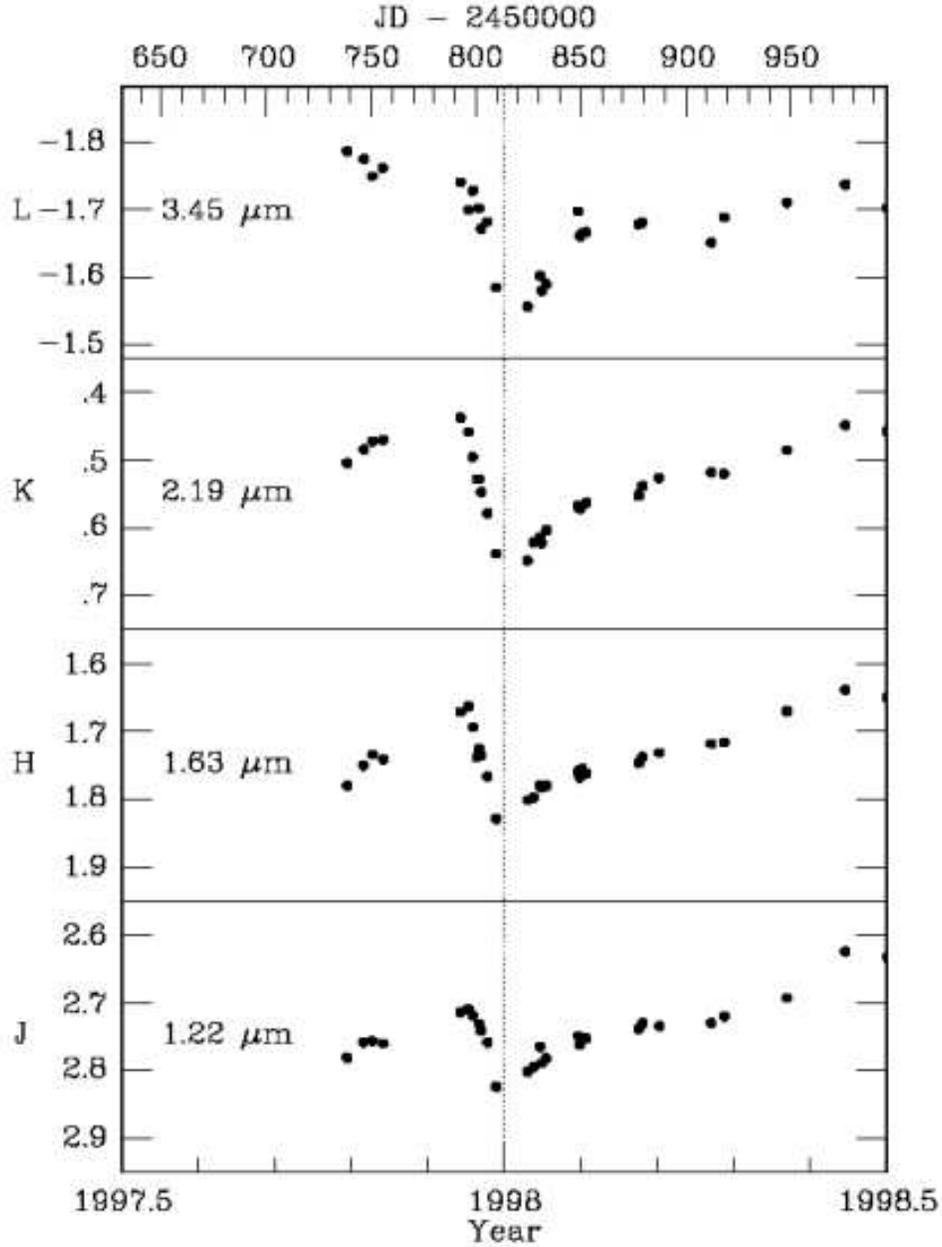}
\caption{Figure 4 from \citet{feast01} reproduced with permission of the authors.  Infrared light curves in $JHKL$ near the time of the 1998 ''dip.''  The minimum of the dip (phase 0.977) is shown by the vertical line.}
\end{figure}

\clearpage

\begin{deluxetable}{llll}
\tablewidth{0pt}
\tablecaption{Brightness of the Central Star at 1999.140}
\tablecolumns{4}
\tablehead{&\colhead{$V_J$}&\colhead{$R_C$}&\colhead{$I_C$}}
\startdata
$\eta$ Car&7.73&6.11&4.47\\
Vega&0.030&0.021&0.025\\
\enddata
\end{deluxetable}

\begin{deluxetable}{lllrrrrrrr}
\tablewidth{0pt}
\tabletypesize{\tiny}
\tablecaption{Results from STIS Acquisition Images}
\tablecolumns{10}
\tablehead{
\colhead{Dataset}&\colhead{MJD}&\colhead{Year}&
\colhead{Flux\tablenotemark{c}}&
\colhead{\% in H$\alpha$\tablenotemark{d}}&
\colhead{Flux @ 6770 \mbox{\AA}\tablenotemark{e}}&
\colhead{Magnitude\tablenotemark{a}}&
\colhead{Average\tablenotemark{b}}&
\colhead{$\sigma$\tablenotemark{b}}&
\colhead{Minus H$\alpha$\tablenotemark{f}}
}
\startdata
o4j802qxq&50814.008&1997.999&3.196E-12&13.2&2.35E-12&0.65&0.65&\nodata&0.79\\
o4j801y4q&50891.414&1998.211&3.720E-12&12.7&3.07E-12&0.49&0.49&\nodata&0.64\\
o55601hkq&51142.172&1998.898&5.374E-12&14.2&4.47E-12&0.09&0.09&\nodata&0.25\\
o55602qsq&51230.449&1999.140&5.831E-12&13.7&5.00E-12&-0.00&-0.00&\nodata&0.16\\
o5f102bkq&51616.484&2000.197&7.476E-12&13.3&\nodata&-0.27&-0.28&0.01&-0.13\\
o5kz01coq&51623.785&2000.217&7.645E-12&\nodata&\nodata&-0.29&\nodata&\nodata&\nodata\\
o5kz02gbq&51624.285&2000.218&7.580E-12&\nodata&\nodata&-0.28&-0.26&0.02&\nodata\\
o5f101yoq&51626.762&2000.225&7.288E-12&\nodata&\nodata&-0.24&\nodata&\nodata&\nodata\\
o5f103a1q&51826.039&2000.769&6.753E-12&\nodata&\nodata&-0.16&-0.16&\nodata&\nodata\\
o62r01h0q&52016.734&2001.291&6.518E-12&11.9&5.60E-12&-0.12&-0.12&\nodata&0.02\\
o6ex03a9q&52183.086&2001.747&7.573E-12&13.5&6.52E-12&-0.28&-0.28&0.00&-0.12\\
o6en01c2q&52183.297&2001.747&7.509E-12&\nodata&\nodata&-0.27&\nodata&\nodata&\nodata\\
o62r02fvq&52240.070&2001.903&7.470E-12&13.5&6.10E-12&-0.27&-0.27&\nodata&-0.11\\
o6ex02tkq&52293.977&2002.051&7.241E-12&14.5&5.47E-12&-0.24&-0.21&0.02&-0.04\\
o6ex01wyq&52294.207&2002.051&6.923E-12&\nodata&\nodata&-0.19&\nodata&\nodata&\nodata\\
o6mo02eaq&52459.480&2002.504&6.933E-12&14.1&5.69E-12&-0.19&-0.20&0.01&-0.04\\
o6mo01k6q&52459.781&2002.505&7.089E-12&\nodata&\nodata&-0.21&\nodata&\nodata&\nodata\\
o8gm01a1q&52624.043&2002.955&6.840E-12&15.5&5.01E-12&-0.17&-0.17&\nodata&0.01\\
o8gm12ukq&52682.859&2003.116&6.730E-12&14.9&5.43E-12&-0.16&-0.12&0.03&0.05\\
o8gm11zpq&52683.242&2003.117&6.326E-12&\nodata&\nodata&-0.09&\nodata&\nodata&\nodata\\
o8gm21t9q&52727.234&2003.238&6.904E-12&13.5&5.29E-12&-0.18&-0.18&\nodata&-0.03\\
o8gm41eqq&52764.266&2003.339&7.462E-12&11.7&5.54E-12&-0.27&-0.27&\nodata&-0.13\\
o8gm33qsq&52776.398&2003.372&7.662E-12&11.2&5.95E-12&-0.30&-0.29&0.00&-0.16\\
o8gm32c6q&52778.469&2003.378&7.623E-12&\nodata&\nodata&-0.29&\nodata&\nodata&\nodata\\
o8gm31fnq&52785.797&2003.398&7.617E-12&10.6&6.24E-12&-0.29&-0.29&0.01&-0.17\\
o8gm51fuq&52791.035&2003.412&7.752E-12&\nodata&\nodata&-0.31&\nodata&\nodata&\nodata\\
o8gm52kyq&52791.707&2003.414&7.558E-12&\nodata&\nodata&-0.28&\nodata&\nodata&\nodata\\
o8gm63ohq&52812.031&2003.470&8.264E-12&9.0&6.98E-12&-0.38&-0.37&0.00&-0.27\\
o8gm61pxq&52812.250&2003.471&8.212E-12&\nodata&\nodata&-0.37&\nodata&\nodata&\nodata\\
o8gm62dwq&52813.703&2003.475&8.222E-12&\nodata&\nodata&-0.37&\nodata&\nodata&\nodata\\
o8ma71jnq&52825.020&2003.506&7.675E-12&8.9&7.05E-12&-0.30&-0.32&0.03&-0.22\\
o8ma72liq&52825.355&2003.506&8.043E-12&\nodata&\nodata&-0.35&\nodata&\nodata&\nodata\\
o8ma81mtq&52849.594&2003.573&7.763E-12&8.3&7.05E-12&-0.31&-0.32&0.01&-0.22\\
o8ma82b8q&52851.906&2003.579&7.850E-12&\nodata&\nodata&-0.32&\nodata&\nodata&\nodata\\
o8ma91ubq&52903.355&2003.720&8.830E-12&\nodata&\nodata&-0.45&-0.48&0.03&\nodata\\
o8ma92ctq&52904.289&2003.723&9.309E-12&\nodata&\nodata&-0.51&\nodata&\nodata&\nodata\\
o8ma83fzq&52960.590&2003.877&9.969E-12&\nodata&\nodata&-0.58&\nodata&\nodata&\nodata\\
\enddata
\tablenotetext{a}{Relative STIS magnitude in the F25ND3 filter zeroed on 1999.140}
\tablenotetext{b}{The average and sigma of individual measurements within one day of each other.  These values are plotted in Figure 4.}
\tablenotetext{c}{Brightness measured in F25ND3 filter given as STIS flux units (erg/cm$^2$/s/\mbox{\AA}).}
\tablenotetext{d}{An estimate of the percent of the total flux measured in the F25ND3 filter which is contributed by H$\alpha$.}
\tablenotetext{e}{The continuum flux measured at 6770 \mbox{\AA} in absolute flux units of erg/cm$^2$/s/\mbox{\AA}.}
\tablenotetext{f}{Relative magnitude in the F25ND3 filter with the contribution from H$\alpha$ subtracted (open triangles in Figure 4).}
\end{deluxetable}

\begin{deluxetable}{lllrrrr}
\tablewidth{0pt}
\tabletypesize{\tiny}
\tablecaption{Results from ACS/HRC Images}
\tablecolumns{7}
\tablehead{
\colhead{Dataset}&\colhead{MJD}&\colhead{Year}&\colhead{Flux\tablenotemark{c}}&
\colhead{Magnitude\tablenotemark{a}}&\colhead{Average\tablenotemark{b}}&
\colhead{$\sigma$\tablenotemark{b}}
}
\startdata
\multicolumn{7}{c}{HRC/F220W Filter}\\
\tableline\\
j8gm1aa7q&52561.039&2002.782&1.385E-12&8.547&8.549&0.011\\
j8gm1aa8q&52561.039&2002.782&1.402E-12&8.533&\nodata&\nodata\\
j8gm1aalq&52561.047&2002.782&1.366E-12&8.562&\nodata&\nodata\\
j8gm1aamq&52561.047&2002.782&1.375E-12&8.554&\nodata&\nodata\\
j8gm1abmq&52561.098&2002.782&1.426E-12&8.515&8.535&0.015\\
j8gm1abnq&52561.098&2002.782&1.404E-12&8.531&\nodata&\nodata\\
j8gm1abyq&52561.105&2002.782&1.394E-12&8.539&\nodata&\nodata\\
j8gm1abxq&52561.105&2002.782&1.374E-12&8.555&\nodata&\nodata\\
j8gm2as1q&52682.547&2003.115&1.562E-12&8.416&8.417&0.007\\
j8gm2as5q&52682.551&2003.115&1.574E-12&8.407&\nodata&\nodata\\
j8gm2asaq&52682.555&2003.115&1.555E-12&8.421&\nodata&\nodata\\
j8gm2asoq&52682.566&2003.115&1.549E-12&8.425&\nodata&\nodata\\
j8ma3adhq&52803.055&2003.445&1.316E-12&8.601&8.571&0.018\\
j8ma3adlq&52803.086&2003.445&1.372E-12&8.557&\nodata&\nodata\\
j8ma3adqq&52803.090&2003.445&1.373E-12&8.556&\nodata&\nodata\\
j8ma3ae3q&52803.152&2003.446&1.355E-12&8.570&\nodata&\nodata\\
j8ma4aqfq&52840.145&2003.547&8.093E-13&9.130&9.129&0.003\\
j8ma4aqj&52840.148&2003.547&8.069E-13&9.133&\nodata&\nodata\\
j8ma4aqnq&52840.156&2003.547&8.109E-13&9.128&\nodata&\nodata\\
j8ma4ar2q&52840.168&2003.547&8.137E-13&9.124&\nodata&\nodata\\
j8ma6ayjq&52957.809&2003.869&9.216E-13&8.989&8.970&0.01\\
j8ma6aynq&52957.844&2003.869&9.531E-13&8.952&\nodata&\nodata\\
j8ma6az2q&52957.855&2003.869&9.385E-13&8.969&\nodata&\nodata\\
\tableline\\
\multicolumn{7}{c}{HRC/F250W Filter}\\
\tableline\\
j8gm1aaaq&52561.039&2002.782&2.935E-12&7.731&7.726&0.011\\
j8gm1aabq&52561.043&2002.782&2.995E-12&7.709&\nodata&\nodata\\
j8gm1aaoq&52561.047&2002.782&2.910E-12&7.740&\nodata&\nodata\\
j8gm1aapq&52561.051&2002.782&2.951E-12&7.725&\nodata&\nodata\\
j8gm1abpq&52561.098&2002.782&2.985E-12&7.713&7.722&0.018\\
j8gm1abqq&52561.098&2002.782&3.022E-12&7.699&\nodata&\nodata\\
j8gm1ac0q&52561.105&2002.782&2.895E-12&7.746&\nodata&\nodata\\
j8gm1ac1q&52561.105&2002.782&2.937E-12&7.730&\nodata&\nodata\\
j8gm2as2q&52682.547&2003.115&3.181E-12&7.643&7.646&0.006\\
j8gm2as6q&52682.551&2003.115&3.198E-12&7.638&\nodata&\nodata\\
j8gm2ascq&52682.559&2003.115&3.174E-12&7.646&\nodata&\nodata\\
j8gm2asqq&52682.570&2003.115&3.148E-12&7.655&\nodata&\nodata\\
j8ma3adiq&52803.055&2003.445&3.058E-12&7.687&7.651&0.021\\
j8ma3admq&52803.086&2003.445&3.194E-12&7.639&\nodata&\nodata\\
j8ma3adsq&52803.094&2003.445&3.184E-12&7.642&\nodata&\nodata\\
j8ma3ae5q&52803.156&2003.446&3.208E-12&7.634&\nodata&\nodata\\
j8ma4aqgq&52840.148&2003.547&2.178E-12&8.055&8.063&0.008\\
j8ma4aqkq&52840.152&2003.547&2.137E-12&8.076&\nodata&\nodata\\
j8ma4aqpq&52840.156&2003.547&2.171E-12&8.058&\nodata&\nodata\\
j8ma4ar4q&52840.168&2003.547&2.159E-12&8.065&\nodata&\nodata\\
j8ma6aqgq&52957.805&2003.869&2.619E-12&7.855&7.840&0.015\\
j8ma6aykq&52957.809&2003.869&2.628E-12&7.851&\nodata&\nodata\\
j8ma6aypq&52957.848&2003.869&2.710E-12&7.817&\nodata&\nodata\\
j8ma6az4q&52957.859&2003.869&2.664E-12&7.836&\nodata&\nodata\\
\tableline\\
\multicolumn{7}{c}{HRC/F330W Filter}\\
\tableline\\
j8gm1aacq&52561.043&2002.782&3.306E-12&7.602&7.621&0.014\\
j8gm1aadq&52561.043&2002.782&3.253E-12&7.619&\nodata&\nodata\\
j8gm1aaqq&52561.051&2002.782&3.243E-12&7.623&\nodata&\nodata\\
j8gm1aarq&52561.051&2002.782&3.189E-12&7.641&\nodata&\nodata\\
j8gm1abrq&52561.098&2002.782&3.378E-12&7.578&7.608&0.021\\
j8gm1absq&52561.102&2002.782&3.292E-12&7.606&\nodata&\nodata\\
j8gm1ac2q&52561.105&2002.782&3.273E-12&7.612&\nodata&\nodata\\
j8gm1ac3q&52561.109&2002.782&3.201E-12&7.637&\nodata&\nodata\\
j8gm2as3q&52682.547&2003.115&3.327E-12&7.595&7.598&0.002\\
j8gm2as7q&52682.551&2003.115&3.312E-12&7.600&\nodata&\nodata\\
j8gm2asgq&52682.563&2003.115&3.313E-12&7.599&\nodata&\nodata\\
j8ma3adjq&52803.055&2003.445&3.913E-12&7.419&7.379&0.023\\
j8ma3adnq&52803.090&2003.445&4.095E-12&7.369&\nodata&\nodata\\
j8ma3adwq&52803.098&2003.445&4.112E-12&7.365&\nodata&\nodata\\
j8ma3ae9q&52803.160&2003.446&4.119E-12&7.363&\nodata&\nodata\\
j8ma4aqhq&52840.148&2003.547&3.671E-12&7.488&7.494&0.003\\
j8ma4aqlq&52840.152&2003.547&3.649E-12&7.495&\nodata&\nodata\\
j8ma4aqtq&52840.160&2003.547&3.648E-12&7.495&\nodata&\nodata\\
j8ma4ar8q&52840.176&2003.547&3.642E-12&7.497&\nodata&\nodata\\
j8ma6aqhq&52957.805&2003.869&4.643E-12&7.233&7.222&0.008\\
j8ma6aylq&52957.809&2003.869&4.695E-12&7.221&\nodata&\nodata\\
j8ma6aytq&52957.852&2003.869&4.738E-12&7.211&\nodata&\nodata\\
j8ma6az8q&52957.852&2003.869&4.678E-12&7.225&\nodata&\nodata\\
\tableline\\
\multicolumn{7}{c}{HRC/F550M Filter}\\
\tableline\\
j8gm1aafq&52561.043&2002.782&3.700E-12&7.479&7.485&0.006\\
j8gm1aatq&52561.051&2002.782&3.663E-12&7.491&\nodata&\nodata\\
j8gm1abuq&52561.102&2002.782&3.755E-12&7.463&7.475&0.014\\
j8gm1ac5q&52561.109&2002.782&3.649E-12&7.494&\nodata&\nodata\\
j8gm1ac4q&52561.109&2002.782&3.745E-12&7.466&\nodata&\nodata\\
j8gm2as4q&52682.551&2003.115&4.129E-12&7.360&7.359&0.004\\
j8gm2as8q&52682.555&2003.115&4.162E-12&7.352&\nodata&\nodata\\
j8gm2asjq&52682.566&2003.115&4.134E-12&7.359&\nodata&\nodata\\
j8gm2asxq&52682.578&2003.115&4.118E-12&7.363&\nodata&\nodata\\
j8ma3adkq&52803.059&2003.445&4.963E-12&7.161&7.126&0.020\\
j8ma3adoq&52803.090&2003.445&5.156E-12&7.119&\nodata&\nodata\\
j8ma3adzq&52803.102&2003.445&5.168E-12&7.117&\nodata&\nodata\\
j8ma3aecq&52803.164&2003.446&5.206E-12&7.109&\nodata&\nodata\\
j8ma4aqiq&52840.148&2003.547&5.259E-12&7.098&7.099&0.003\\
j8ma4aqmq&52840.152&2003.547&5.250E-12&7.100&\nodata&\nodata\\
j8ma4aqwq&52840.164&2003.547&5.234E-12&7.103&\nodata&\nodata\\
j8ma4arbq&52840.180&2003.547&5.272E-12&7.095&\nodata&\nodata\\
j8ma6ayiq&52957.805&2003.869&6.851E-12&6.811&6.807&0.004\\
j8ma6aymq&52957.813&2003.869&6.907E-12&6.802&\nodata&\nodata\\
j8ma6aywq&52957.855&2003.869&6.890E-12&6.804&\nodata&\nodata\\
j8ma6azbq&52957.867&2003.869&6.849E-12&6.811&\nodata&\nodata\\
\enddata
\tablenotetext{a}{Magnitude on the STMAG system.}
\tablenotetext{b}{The average and sigma of individual measurements in an exposure set.  These values are plotted in Figure 6.}
\tablenotetext{c}{STMAG flux units are erg/cm$^2$/s/\mbox{\AA}.}
\end{deluxetable}

\end{document}